\documentclass{ws-p8-50x6-00}

\begin{document}

\title{The Quasi-Molecular Stage of Ternary Fission}

\author{D. N. Poenaru and  B. Dobrescu}

\address{Horia Hulubei National Institute of Physics and Nuclear Engineering, 
\\P.O. Box MG-6, RO-76900 Bucharest, Romania\\E-mail: poenaru@ifin.nipne.ro}

\author{W. Greiner}

\address{Institut f\"ur Theoretische Physik der Universit\"at, Postfach
111932, \\D-60054 Frankfurt am Main, Germany} 

\author{J. H. Hamilton and A. V. Ramayya}

\address{Department of Physics, Vanderbilt University, Nashville,
Tennessee, USA}

\maketitle

\abstracts{We developed a three-center phenomenological model,
able to explain
qualitatively the recently obtained experimental results concerning the
quasimolecular stage of a light-particle accompanied fission process.
It was derived from the liquid drop model under the assumption 
that the aligned  configuration, with the emitted particle between the
light and heavy fragment, is reached by increasing continuously
the separation distance, while the radii of the heavy fragment and of the
light particle are kept constant. In such a way, 
a new minimum of a
short-lived molecular state appears in the deformation energy at a 
separation distance very close to the touching point. 
This minimum 
allows the existence of a short-lived quasi-molecular state, decaying 
into the three final fragments.
The influence of the shell effects is discussed. The half-lives of some
quasimolecular states which could be formed in the $^{10}$Be
and $^{12}$C accompanied 
fission of $^{252}$Cf are roughly estimated to be the order of 1~ns, and
1~ms, respectively.
}

\section{Introduction}

The light particle accompanied fission was
discovered~\cite{san46cr} in 1946, when the track of a long-range 
particle (identified by Farwell {\em et al.} to be $^4$He)
almost perpendicular to the short tracks of heavy and light fragments was
observed in a photographic plate. The fission was induced by
bombarding $^{235}$U with slow neutrons from a Be target at a cyclotron.
The largest yield in such a rare process (less than one event per 500
binary splittings) is measured for $\alpha $-particles, but many other light
nuclei (from protons to oxygen or even calcium isotopes) have been 
identified~\cite{mut96mb} in both induced- and spontaneous fission phenomena. 
If $A_1$ and $A_2$ are the mass numbers of the heavy fragments (assume 
$A_1 \geq A_2$),
then usually the mass of the light particle $A_3 < < A_2$. The ``true''
ternary fission, in which $A_1 \simeq A_2 \simeq A_3$, has not yet been
experimentally detected. 

Many properties of the binary fission process have been 
explained~\cite{boh39pr} within the liquid drop model (LDM); 
others like the
asymmetric mass distribution of fragments and the ground state deformations
of many nuclei, could be understood 
only  after adding the contribution of shell
effects.~\cite{str67np,mye66np} As it was repeatedly stressed 
(see~\cite{p195b96b} and the references therein),  shell effects proved 
also to be of vital importance for cluster radioactivities 
predicted~\cite{ps84sjpn80} in 1980.
\begin{figure}
\centerline{\epsfxsize=1.5in\epsffile{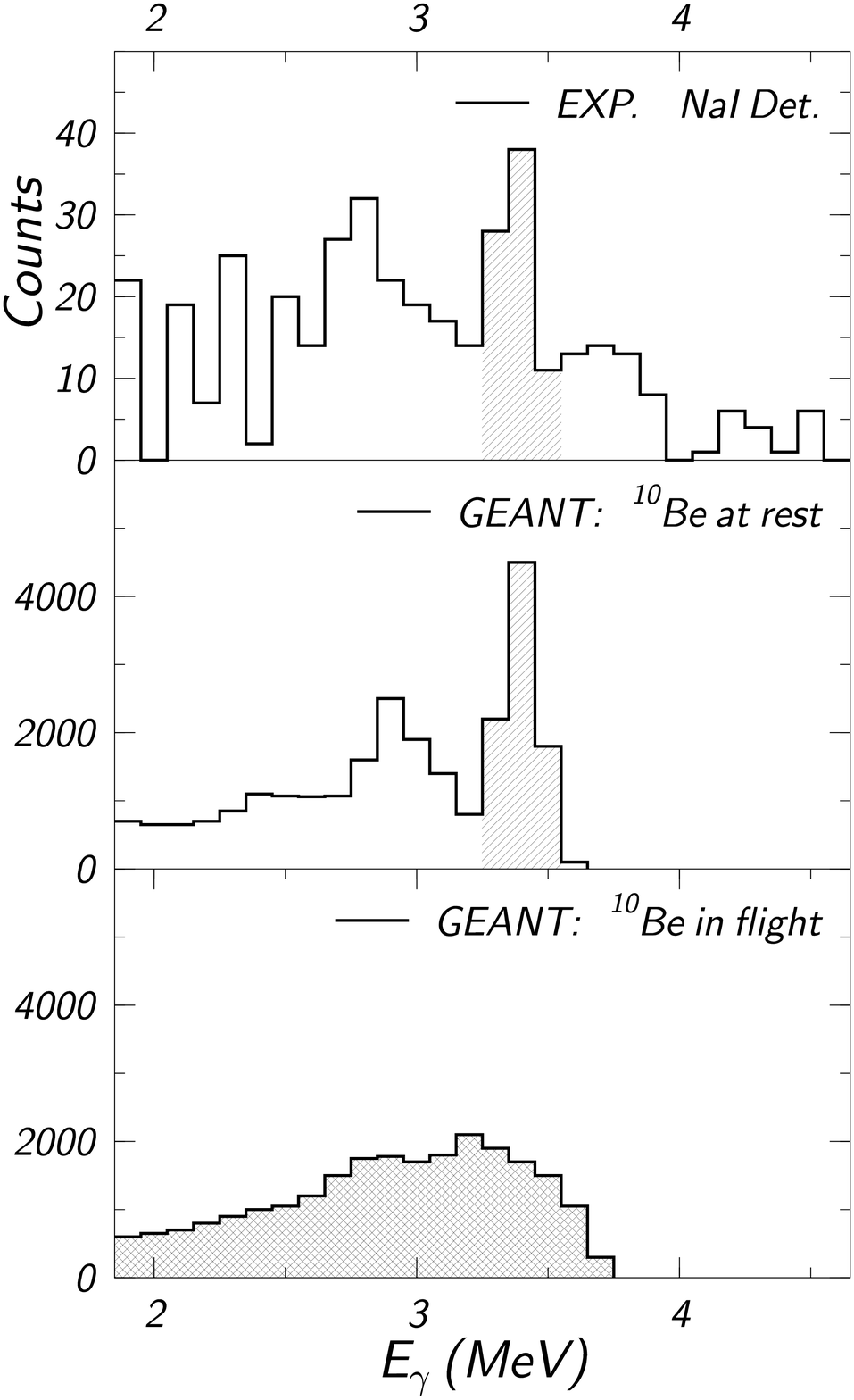}}
\bigskip
\caption{Gamma-ray energy spectrum of $^{10}$Be accompanying fission of
$^{252}$Cf (top) and the simulated (with the code GEANT) pulse-height
distributions for a (not-Doppler-broadened) 3.37~MeV $\gamma$~emission 
from $^{10}$Be at rest (middle) as well as for $^{10}$Be in flight (bottom)
from the
source to the detector (duration of about 1~ns). The flight-path of the
order of 10~cm. One uses NaI scintillator detectors.
Data from Fig.~6 of Ref. 13.}
\label{fig:mut}
\vspace{-4mm}
\end{figure}

The total kinetic energy (TKE) of the fragments, in the most frequently
detected binary or ternary fission mechanism, is smaller than the released
energy ($Q$) by about 25--35~MeV, which is used to produce deformed and
excited fragments. These then emitt neutrons (each with a binding energy 
of about 6~MeV)
and $\gamma$-rays. From time to time a ``cold'' fission mechanism is
detected, in which the TKE exhausts the $Q$-value, 
hence no neutrons are emitted,
and the fragments are produced in or near 
their ground-state. The first experimental
evidence for cold binary fission 
in which its TKE exhaust $Q$ was reported~\cite{sig81jpl} in 1981. 
Larger yields were measured~\cite{hul86} in trans-Fm ($Z\geq 100$) 
isotopes, where the phenomenon was called bimodal fission.
\begin{figure}
\centerline{\epsfxsize=1.75in\epsffile{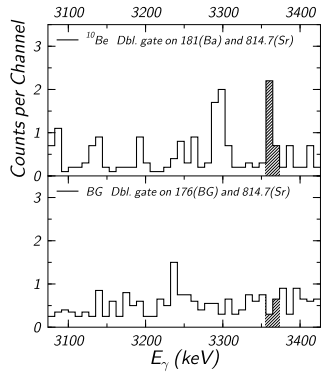}}
\bigskip
\caption{The cumulated $\gamma$-ray energy spectrum of $^{10}$Be 
accompanying fission of
$^{252}$Cf (top) and the corresponding background (bottom).
The spectrum of the 3.362 $\gamma$ is not-Doppler-broadened, suggesting an
emission from $^{10}$Be at rest. The stopping time of $^{10}$Be in the
absorber mounted around the source is of the order of 1~ps.
Data from Fig.~2 of Ref. 11. 
Coincidence spectra obtained with Ge detectors in GAMMASPHERE.}
\label{fig:ram}
\vspace{-4mm}
\end{figure}

The correlated fragment pairs in
cold ternary ($\alpha$- and $^{10}$Be accompanied spontaneous fission of
$^{252}$Cf) processes were only recently discovered,~\cite{ram98pr,ram98prl}
by measuring triple~$\gamma $ coincidences in a modern large
array of $\gamma$-ray detectors (GAMMASPHERE). The fragments 
are identified by their $\gamma$-ray spectra. Among other new aspects of the
fission process seen for the first time  with this new 
technique,~\cite{ram98pr,ham95ppnp} 
one should mention the double fine
structure, and the triple fine structure in binary and ternary fission.

A particularly interesting feature, observed~\cite{ram98prl,sin96cp}
both in $^{10}$Be- and $^{12}$C accompanied cold fission of $^{252}$Cf 
is related to the width of the light particle $\gamma$-ray spectrum (see
Figs.~\ref{fig:mut} and \ref{fig:ram}).
For example, the 3.368~MeV $\gamma$ line of $^{10}$Be, 
with a lifetime of 125~fs is not Doppler-broadened,
as it should be if it would be emitted when $^{10}$Be is 
in flight (taking about 1~ns to reach the detector).
A plausible suggestion 
was made, that the absence of Doppler broadening is related to
a trapping of $^{10}$Be in a potential well 
of nuclear molecular character.~\cite{ram98prl}

Quasi-molecular configurations of two nuclei have been suggested as a 
natural explanation for the resonances measured~\cite{bro60prl}
in $^{12}$C$+^{12}$C scattering and reactions. There are also other kinds
of such binary molecules (see~\cite{gre95b} and references
therein), like spontaneously fissioning shape-isomers. The above 
mentioned experiments can be considered as the first evidence for a 
more complex quasi-molecular configuration of three nuclei.
The purpose of the present lecture is to show, within a 
phenomenological three-center model, that a
minimum which could explain the existence of 
these quasi-molecules is produced in the potential barrier, 
when the formation of the light particle occurs in 
the neck between the two heavier fragments.
In this way we extend to ternary fission our unified approach of cold
fission, cluster radioactivities, and $\alpha$-decay.~\cite{p195b96b}

\section{Shape Parametrization}

\begin{figure}[b]
\centerline{\epsfxsize=2in\epsffile{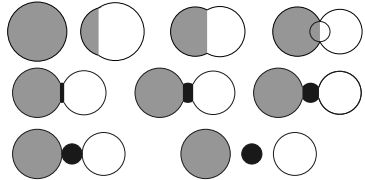}}
\bigskip
\caption{Evolution of nuclear shapes during the deformation process from
one parent nucleus $^{252}$Cf to three separated fragments $^{146}$Ba,
$^{10}$Be, and $^{96}$Sr. In the upper part the binary stage is illustrated;
the separation distance increases from $R_i$ to $R_{ov3}$, passing through
$R_{min1b}$ and $R_{min2b}$ values. In the middle, the ternary stage of the
process develops by forming the third particle in the neck. The
quasi-molecular shape, at which $R=R_{min-t}$ is the intermediate 
one in this row. At the bottom the fragments are separated.}
\label{fig:shapes}
\vspace{-4mm}
\end{figure}
The shape parametrization with one deformation parameter as follows has 
been suggested from the analysis~\cite{p218pr99} of different aligned and 
compact configurations of fragments in touch. A lower
potential barrier for the aligned cylindrically-symmetric shapes
with the light particle between the two heavy fragments, is a clear 
indication that during the deformation from an initial parent nucleus to
three final nuclei, one should arrive at such a scission point.
In order to reach this stage we shall increase continuously
the separation distance, $R$, between the heavy fragments, while the 
radii of the heavy fragment and of the
light particle are kept constant, $R_1=$~constant, $R_3=$~constant. 
Unlike in the previous work, we now adopt the following convention:
$A_1 \geq A_2 \geq A_3$. 
The hadron numbers are conserved: $A_1+A_2+A_3=A$.

At the beginning (the neck radius $\rho_{neck} \geq R_3$) one has 
a two-center evolution (see Fig.~\ref{fig:shapes}) 
until the neck between the fragments becomes 
equal to the radius of the emitted particle, 
$\rho_{neck}=\rho(z_{s1})\mid _{R=R_{ov3}}=R_3$.  This Eq. defines $R_{ov3}$
as the separation distance at which the neck radius is equal to $R_3$.
By placing the origin in the center of the large sphere, the surface 
equation in cylindrical coordinates is given by:
\begin{equation}
\rho_s^2 = \left \{ \begin{array}{lll} 
 &R_1^2-z^2 & \;\;\;,\;\; \mbox{$-R_1\leq z\leq z_{s1}$} \\ 
 &R^2_2-(z-R)^2 & \;\;\;,\;\; \mbox{$z_{s1}\leq z\leq R+R_2$} 
\end{array} \right.
\end{equation}
Then for $R > R_{ov3}$ the three center starts developing by
decreasing progressively with the same amount the two tip distances
$h_1+h_{31}=h_{32}+ h_2$. Besides this constraint, one has as in the binary 
stage, volume conservation and matching conditions. The $R_2$ and the other
geometrical quantities are determined by solving numerically the
corresponding system of
algebraic equations. By assuming spherical nuclei, the radii are given by
$R_j=1.2249A_j^{1/3}$~fm ($j=0, 1, 3$), $R_{2f}=1.2249A_2^{1/3}$ with a
radius constant $r_0=1.2249$~fm, from Myers-Swiatecki's variant of LDM.
Now the surface equation can be written as
\begin{equation}
\rho_s^2 = \left \{ \begin{array}{lll} 
 &R_1^2-z^2 & \;\;\;,\;\; \mbox{$-R_1\leq z\leq z_{s1}$} \\ 
 &R^2_3-(z-z_3)^2 & \;\;\;,\;\; \mbox{$z_{s1}\leq z\leq z_{s2}$} \\ 
&R^2_2-(z-R)^2 & \;\;\;,\;\; \mbox{$z_{s2}\leq z\leq R+R_2$} 
\end{array} \right. 
\end{equation}
and the corresponding shape has two necks and two separating planes.
Some of the important values of the deformation parameter $R$ are the
initial distance $R_i=R_0-R_1$, and the touching-point one, 
$R_t=R_1 + 2R_3 + R_{2f}$. There is also $R_{ov3}$, defined above, 
which allows one to distinguish between the binary and ternary stage.

\section{Deformation Energy}

According to the LDM, by requesting zero energy for 
a spherical shape, the deformation energy, $E^u(R) - E^0$,  is expressed 
as a sum of the surface and Coulomb terms
\begin{equation}
E_{def}^u(R) = 
E_s^0[B_s(R) - 1] + E_C^0[B_C(R) -1]
\end{equation}
where the exponent $^u$ stands for uniform (fragments with the same charge
density as the parent nucleus), and $^0$ refers to the initial spherical
parent. In order to
simplify the calculations, we initially assume the same charge density
$\rho _{1e}=\rho _{2e}=\rho _{3e}=\rho _{0e}$, and at the end we add the
corresponding corrections.  In this way we perform one
numerical quadrature instead of six.
For a spherical shape 
$E_s^0=a_s(1-\kappa I^2)A^{2/3}$ ; \ \ $I=(N-Z)/A$; 
$E_C^0 = a_cZ^2A^{-1/3}$, where the numerical constants of the LDM are:
$a_s=17.9439$~MeV, $\kappa =1.7826$, $a_c=3e^2/(5r_0)$, 
$e^2=1.44$~MeV$\cdot$fm.

The shape-dependent, dimensionless surface term is proportional to the
surface area:
\begin{equation}
B_{s}=\frac{E_s}{E_s ^0} = 
\frac{d^2}{2} \int\limits_{-1}^{+1} \left [y^2+\frac{1}{4}
\left ( \frac{dy^2}{dx}\right ) ^2 \right ]^{1/2} dx 
\end{equation}
where $y = y(x)$ 
is the surface equation in cylindrical 
coordinates with -1, +1 intercepts on the symmetry axis, and
$d = (z'' - z')/2R_0$ is the seminuclear length in units of $R_0$.
Similarly, for the Coulomb energy~\cite{p75cpc78} one has
\begin{equation}
B_{c}=\frac{5d^5}{8\pi} \int\limits_{-1} ^{+1} dx \int\limits_{-1} ^{+1} 
dx' F(x, x') 
\end{equation}
\begin{eqnarray} 
F(x,x')&=&\{ y y_1[(K-2D)/3]\cdot \nonumber \\ & &
\nonumber \left [ 2(y^2+y_1^2)-(x-x')^2+\frac{3}{2}(x
-x')\left ( \frac{dy_1^2}{dx'}-\frac{dy^2}{dx} \right ) \right ]+
\nonumber \\
 & &K \left \{ y^2y_1^2/3+\left [y^2-\frac{x-x'}{2}\frac{dy^2}{dx}
\right ] \left [y_1^2-\frac{x-x'}{2}\frac{dy_1^2}{dx'}\right ] 
\right \} \} a_{\rho}^{-1} 
\end{eqnarray} 
$K$, $K'$ are the complete elliptic integrals of the 1st and 2nd kind
\begin{equation}
K(k) = \int\limits_0^{\pi /2}(1-k^2 {\sin}^2 t)^{-1/2} dt ; \ \ 
K'(k) = \int\limits_0^{\pi /2}(1-k^2 {\sin}^2 t)^{1/2} dt 
\end{equation}
and $a_{\rho} ^2 = (y+y_1)^2+(x-x')^2$, $k^2 = 4yy_1 /a_{\rho}^2$, $D
= (K - K')/k^2$.
\begin{figure}
\centerline{\epsfxsize=2.5in\epsffile{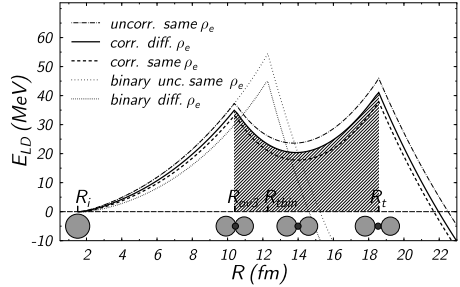}}
\caption{The liquid drop model deformation energy versus separation distance
for the $^{20}$O accompanied cold fission of $^{252}$Cf with $^{132}$Sn 
and $^{100}$Zr heavy fragments. 
In order to simplify the numerical calculations we start by
assuming the same charge density of the fragments.
One can see the 
effect of successive corrections taking into account the
experimental $Q$-value and the difference in charge density.
Similar curves for the binary fission posses a narrower fission
barrier.  The new minimum appears in the shaded area from $R_{ov3}$
to $R_t$.}
\label{fig:ldm20o}
\vspace{-4mm}
\end{figure}

The new minimum, which can be seen in Fig.~\ref{fig:ldm20o} 
at a separation distance
$R=R_{min-t}>R_{ov3}$, 
is the result of a competition between the Coulomb- and surface
energies. At the beginning ($R<R_{min-t}$) 
the Coulomb term is stronger, leading to a
decrease in energy, but later on ($R>R_{min-t}$) 
the light particle formed in the neck posses a surface area increasing 
rapidly, so there is also an increase in energy up to $R=R_t$.

Now let us analyse the influence of various corrections, which could in
principle alter this image.
After performing numerically the integrations, we add the following 
corrections: for the difference in charge densities reproducing the touching 
point values; for experimental masses reproducing the $Q_{exp}$-value at 
$R=R_i$, when the origin of energy corresponds to infinite separation
distances between fragments, 
and the phenomenological shell corrections $\delta E$
\begin{equation}
 E_{LD}(R)= E_{def}^u(R) + (Q_{th}-Q_{exp})f_c(R) 
\end{equation}
where $f_c(R)=(R-R_i)/(R_t-R_i)$, and
\begin{equation}
Q_{th} 
=E^0_s+E^0_C-\sum_1^3(E^0_{si}+E^0_{Ci})
+\delta E^0 - \sum_1^3\delta E^i
\end{equation}
The correction increases gradually (see Fig.~\ref{fig:ldm20o} and
Fig.~\ref{fig:ld+sbe})
with $R$ up to $R_{t}$ and then 
remains constant for $R>R_t$. The barrier height increases if $Q_{exp}<Q_{th}$ 
and decreases if $Q_{exp}>Q_{th}$. In this way, when one, two, or all final
nuclei have magic numbers of nucleons, $Q_{exp}$ is large and the fission
barrier has a lower height, leading to an increased yield. In a binary 
decay mode like cluster radioactivity and cold fission, this condition 
is fulfilled when the daughter nucleus is $^{208}$Pb and $^{132}$Sn,
respectively.

\section{Shell Corrections and Half-lives}

\begin{figure}
\centerline{\epsfxsize=2.1in\epsffile{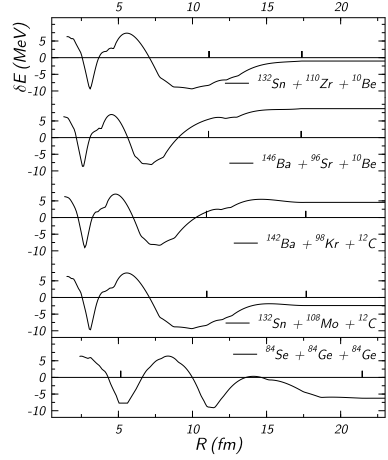}}
\caption{ The shell correction energy variation with the separation distance
for two examples of $^{10}$Be-, two of $^{12}$C accompanied cold fission,
compared to the ``true'' ternary fission (in nearly three identical
fragments) of $^{252}$Cf. The three partners are given. The
two vertical bars on each plot show the positions of $R_{ov3}$ and of 
$R_t$.}
\label{fig:shell}
\vspace*{-0.6cm}
\end{figure}

Finally we also add the shell terms
\begin{equation}
E(R)=E_{LD}(R) + \delta E(R) - \delta E^0
\end{equation}
\begin{figure}
\centerline{\epsfxsize=2.5in\epsffile{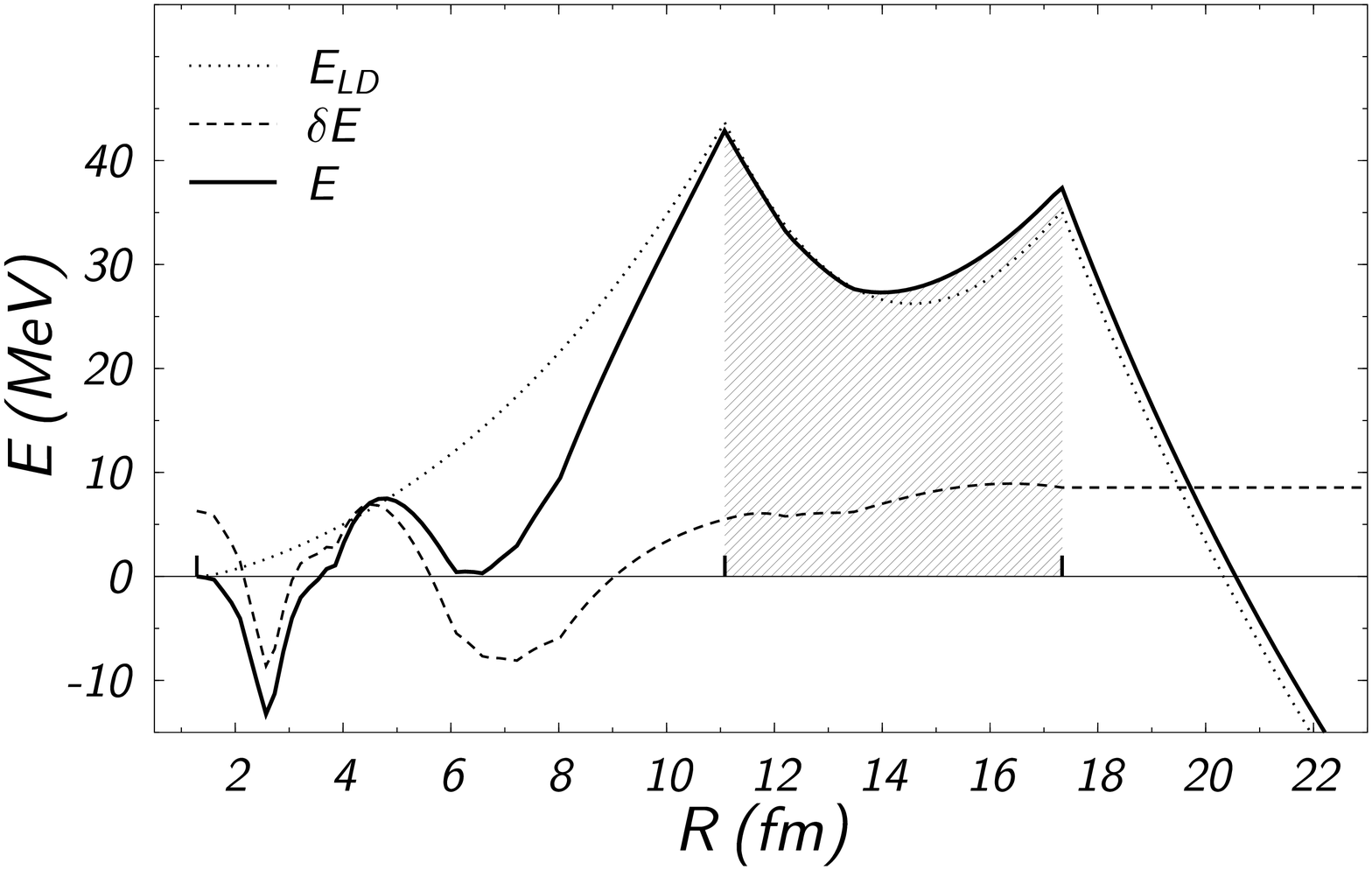}}
\caption{ The liquid drop model, $E_{LD}$, the shell correction, $\delta E$,
and the total deformation energies, $E$, for the $^{10}$Be accompanied 
cold fission of $^{252}$Cf with $^{146}$Ba and $^{96}$Sr heavy
fragments. The new minimum appears in the shaded area from $R_{ov3}$
to $R_t$.}
\label{fig:ld+sbe} 
\vspace{-4mm}
\end{figure}
Presently there is not available any microscopic 
three-center shell model reliably
working for a long range of mass asymmetries. 
This is why we use a phenomenological model,
instead of the Strutinsky's method, to calculate the shell corrections.
The model is adapted after Myers and Swiatecki.~\cite{mye66np}
At a given $R$, we calculate the volumes of fragments and the corresponding
numbers of nucleons $Z_i(R), \ N_i(R)$ ($i=1,2,3$), 
proportional to the volume of each
fragment. Fig.~\ref{fig:shapes} illustrates the evolution of shapes and of the 
fragment volumes. Then we add for each fragment the contribution of protons
and neutrons
\begin{equation}
\delta E(R) = \sum_i \delta E_i(R) = \sum_i [\delta E_{pi}(R) + \delta
E_{ni}(R)] 
\end{equation}
which are given by
\begin{equation}
 \delta E_{pi}=Cs(Z_i) ; \ \ \delta E_{ni}=Cs(N_i)
\end{equation}
where
\begin{equation}
 s(Z) = F(Z)/[(Z)^{-2/3}] -cZ^{1/3} 
\end{equation}
\begin{equation}
F(n) = \frac{3}{5}\left [\frac{N_i^{5/3} -N_{i-1}^{5/3}}{N_i -
N_{i-1}}(n -N_{i-1}) - n^{5/3}+ N_{i-1}^{5/3} \right ]
\end{equation}
in which $n \in (N_{i-1}, N_i)$ is either a current 
$Z$ or $N$ number and 
$N_{i-1}, N_i$ are the closest magic numbers. The constants 
$c=0.2$, $C=6.2$~MeV were 
determined by fit to the experimental masses and deformations.
The variation with $R$ is calculated~\cite{sch71pl} as
\begin{equation}
\delta E(R) = \frac{C}{2}\left \{ \sum_i[s(N_i)+s(Z_i)]\frac{L_i(R)}{R_i} 
\right \}
\end{equation}
where $L_i(R)$ are the lengths of the fragments along the axis of symmetry, 
at a given separation distance $R$.
During the deformation, the variation of separation distance between
centers, $R$, induces the variation of the geometrical quantities and of the
corresponding nucleon numbers. Each
time a proton or neutron number 
reaches a magic value, the correction energy
passes through a minimum, and it has a maximum at midshell 
(see Fig.~\ref{fig:ld+sbe} and Fig.~\ref{fig:shell}). 
The first narrow minimum appearing in the shell correction energy
$\delta E$ in Fig.~\ref{fig:ld+sbe}, at $R=R_{min1b}\simeq 2.6$~fm, is the
result of almost simultaneously reaching the magic numbers $Z_1=20$,
$N_1=28$, and $Z_2=82$, $N_2=126$. The second, more 
shallower one around
$R=R_{min2b}\simeq 7.2$~fm corresponds to a larger range of $R$-values for
which $Z_1=50$, $N_1=82$, $Z_2=50$, $N_2=82$ are not obtained in the same
time. In the region of the new minimum, $R=R_{min-t}$, for light-particle
accompanied fission, the variation of the shell correction energy is 
very small, hence it has no major consequence. One can say that the
quasimolecular minimum is related
to the collective properties (liquid-drop like
behavior). On the other side, for
``true'' ternary process (see the bottom part of Fig.~\ref{fig:shell}) 
both minima appear
in this range of values, but no such LDM effect was found there.
In order to compute the half-life of the quasi-molecular state, we have
first to search for the minimum $E_{min}$ in the 
quasimolecular well, from $R_{ov3}$ to $R_t$, and then to add a zero point
vibration energy, $E_v$: $E_{qs}=E_{min} + E_v$.


The half-life, $T$, is expressed in terms of the barrier penetrability, $P$, 
which is calculated from an action integral, $K$, given by the 
quasi-classical WKB approximation
\begin{equation}
 T=\frac{h \ln 2 }{2 E_v P} ; \ \ P=exp(-K) 
\end{equation}
where $h$ is the Planck constant,  and 
\begin{equation}
K=\frac{2}{\hbar} \int\limits_{R_a}^{R_b} 
\sqrt{2\mu[E(R)-E_{qs}]} \ dR 
\end{equation}
in which $R_a, R_b$ are the turning points, defined by 
$E(R_a)=E(R_b)=E_{qs}$
and the nuclear inertia is roughly approximated  by the reduced mass 
$\mu=m[(A_1A_2+A_3A)/(A_1+A_2)]$, where $m$ is the nucleon mass,
$\log[(h \ln 2)/2]=-20.8436$, \ \ $\log e = 0.43429$ and 
$\sqrt{8m/\hbar^2}=0.4392$~MeV$^{-1/2}\times$fm$^{-1}$. 

The results of our estimations for the half-lives of some
quasimolecular states formed in the $^{10}$Be- and $^{12}$C accompanied
fission of $^{252}$Cf are given in Table~1. 
They are of the order of 1~ns and 1~ms, respectively, if we ignore the
results for a division with heavy fragment $^{132}$Sn, which was not
measured due to very high first excited state. Consequently
the new minimum we found can qualitatively
explain the quasimolecular nature of the narrow line of the $^{10}$Be
$\gamma $-rays.

\begin{table}[t]
\caption{Calculated half-lives of some quasi-molecular states formed during
the ternary fission of $^{252}$Cf.}
\begin{center}
\begin{tabular}{|c|cc|c|c|r|}
\hline
{Particle}&\multicolumn{2}{|c|}{Fragments} & {$Q_{exp}$}&{$K$}& $\log T(s)$\\
 & & & (MeV) &  & \\
\hline
 & & & & & \\
$^{10}$Be & $^{132}$Sn & $^{110}$Ru & 220.183 &  19.96& -11.17 \\
          & $^{138}$Te & $^{104}$Mo & 209.682 &  25.23& -8.89 \\
          & $^{138}$Xe & $^{104}$Zr & 209.882 &  26.04& -8.54 \\
          & $^{146}$Ba & $^{96}$Sr  & 201.486 &  22.98& -9.86 \\ 
 & & & & & \\
$^{12}$C & $^{147}$La & $^{93}$Br   & 196.268 &  39.80& -2.56 \\
         & $^{142}$Ba & $^{98}$Kr   & 199.896 &  42.71& -1.30 \\
         & $^{140}$Te & $^{100}$Zr  & 209.728 &  38.21& -3.25 \\
         & $^{132}$Sn & $^{108}$Mo  & 223.839 &  31.46& -6.18 \\
\hline  
\end{tabular}
\end{center}
\end{table}

It is interesting to note that the trend toward a split into two, three, or
four nuclei (the lighter ones formed in a long neck between the heavier
fragments) has been theoretically demonstrated by Hill,~\cite{hil58c}
who
investigated the classical dynamics of an incompressible, irrotational,
uniformly charged liquid drop.
No mass asymmetry was evidenced since any shell effect was ignored.

In conclusion, we should stress that a quasimolecular stage of a
light-particle accompanied fission process, for a limited range of sizes of
the three partners, can be qualitatively
explained within the liquid drop model.



\subsection*{Acknowledgments}
We are grateful to M. Mutterer for enlightening discussions.


\end{document}